# Carrier-potential interaction for high-$T_c$ superconductivity

Hui Pan[†+]

[†]Institute of Applied Physics and Materials Engineering, University of Macau, Macao SAR

[+]Department of Physics and Chemistry, Faculty of Science and Technology, University of Macau, Macao SAR

**Abstract:** The origin of high-temperature superconductivity has been widely debated since its discovery. Here, a model based on the interaction between carrier and local potential is proposed to reveal the mechanism. In this model, the potential that is analogous to the lattice point is composed of localized charges, whose vibration mediates the coupling of mobile carriers. A Hamiltonian that describes the vibration, coupling, and various interactions among the ordered potentials and carriers is established. By analyzing the Hamiltonian, it is found that the vibration of local potential and the interactions, which are dependent on the carrier density, control the transition temperature. We show that the transition temperature is high if the local potential is composed of electrons and the mobile carrier is hole because of the strong coupling between them. By replacing the local potential with lattice point, the proposed model is equivalent to that in the BCS theory. Therefore, our model may provide a general theoretical description on the superconductivity.

Key Words: Ordered potentials; carrier-potential interaction; high-temperature superconductivity; Hamiltonian

PACS: 74.20.-z;

Hui Pan: huipan@um.edu.mo (email), 853-88224427 (tel.), 853-88222454 (fax).

## I. Introduction

Superconductivity is an amazing topic that has been attracting great interests from physicists and materials scientists over decades [1-8]. Extensive studies have been conducted, experimentally and theoretically, to reveal the mechanism that triggered the phenomenon [9-19]. It is well known that there are two types of superconductors i.e., low-temperature (conventional) and high-temperature (unconventional) superconductors, in view of the transition temperature ($T_c$). In 1957, Bardeen, Cooper, and Schrieffer presented that the 'Cooper' pairs are formed due to the weak attraction between electrons that is induced by the vibration of lattice point (phonon), as bosons condense into a superfluid state, which successfully provides an understanding to the conventional superconductor (the BCS theory) [9]. Although this role is capable of producing $T_c$ up to 30 K, it is difficult to explain the superconductivity with $T_c$ above 90 K as observed in copper oxides [4].

High-temperature (high-$T_c$) superconductivity (HTS) can be achieved by doping Mott-insulating copper oxides with electrons or holes [3, 4], where the transition temperature is strongly dependent on the density and types of carriers [13]. Numerous experimental works have suggested that the charge-charge interaction plays a significant role in the high $T_c$ superconductivity [10-19]. To describe the strong collective-charge interaction, there are many efforts have been conducted. The resonating-valence-bond theory (RVB) proposed by Philip Anderson stated that neighboring copper atoms are linked through chemical valence bond and share electrons with opposite spins [10]. Without doping, these spin pairs are locked in places, leading to Mott-insulator. When doped with electrons or holes, the valence bonds become Cooper pairs and mobile, resulting in superconductor. Another important theory is spin fluctuation, which was devised by Philippe Monthoux, Alexander Balatsky, and David Pines [19]. Without doping, the spins are locked into an ordered state of an antiferromagnet, while they are fluctuated in doped cuprates, where the fluctuation draws moving electrons together to form the Cooper pairs and result in superconducting state. However, the nature and mechanism behind the high-temperature superconductivity have not been yet identified conclusively to explain everything, such as pseudogap and strange metal behavior [22].

It is now widely accepted that the collective charge dynamics should be responsible for the observed phenomena [23-26]. For example, Hepting et al. recently reported that the 'acoustic plasmon', a branch of distinct charge collective modes, played a substantial part in mediating high-temperature superconductivity [26]. Both RVB theory and Spin Fluctuation described the charge dynamics in different ways. In general, the mediator for the Cooper pairs in superconductor can be viewed as the vibration of potential, that is, the phonon in the BCS theory, the RVB in Anderson's theory, and the spin fluctuation from Monthoux, et al.

**II. Hamiltonian**

Analogous to the lattice vibrations in conventional superconductivity as provided in the BCS theory, in this work, we propose that the pulsation of ordered local-potentials mediates the paring of carriers, which results in superconductivity. Firstly, the potential is the localized charges in this model, which may be composed of a few of electrons or holes. For example, the potential may be dominated by two electrons with opposite spins.

Different from the ions in crystal, all the potentials have the same charge state, positive or negative. Secondly, we consider that the potential is movable (local or dispersive) due to the interaction with carriers, such as carrier annihilation and creation. Therefore, the potential itself may be one kind of Cooper pairs in high-T$_c$ superconductor. Finally, the charge state of potential is opposite to that of carrier, leading to the attraction interaction, and the annihilation and creation of potentials and carriers as induced by the coupling among them. At the same time, there have interactions among potentials.

According to the above assumption on the interaction between potential and carrier, the Hamiltonian of the HTS system is obtained as below if we consider that the potential is composed of two electrons or holes with opposite spins:

$$H = \varepsilon_a \sum_i a_i^+ a_i + \sum_{<i,j>} v_{ij}^a n_i^a n_j^a + \varepsilon_b \sum_k b_i^+ b_i + \sum_{<k,l>} v_{kl}^c n_k^c n_l^c + \sum_{<i,k>} v_{ik}^{ac} n_i^a n_k^c + t_{ab} \sum_{<i,k>} (a_i^+ b_k^+ + hc.)$$

$$+ t_{dc} \sum_{<i,j,k>} (a_i d_j^+ c_k + hc.) \qquad (1)$$

$$a_i^+ = d_i^+ d_{-i}^+, a_i = d_i d_{-i}$$

$$b_k^+ = c_k^+ c_{-k}^+, b_k = c_k c_{-k}$$

$$n_i^a = a_i^+ a_i, n_k^c = c_k^+ c_k$$

where $d_i^+$ and $d_i$ are the operators of creation and annihilation of charge (electron or hole) in potential, $a_i^+$ and $a_i$ are the operators of creation and annihilation of the pair of electrons or holes in potential, $c_k^+$ and $c_k$ are the operators for creating and annihilating carriers, and $b_k^+$ and $b_k$ are the operators for creating and annihilating the pair of carriers. The creation of charge in the potential is opposite to that of carrier. $d_i^+$ is the creation of electron if $c_k^+$ is for the creation of hole, and vise verse. The first two terms describe the total energy of potentials, where $\varepsilon_a$ and $v_{ij}^a$ are the kinetic and interaction energies, respectively. The third and fourth terms are the total energy of carriers, where $\varepsilon_b$ and $v^b_{kl}$ are the kinetic and interaction energies, respectively. The fifth term is the interaction between potential and carrier, where $v^{ac}_{ij}$ is the interaction energy. The last two terms are used to describe the recombination between the potential and carrier, where $t_{ab}$ and $t_{dc}$ are the full and partial recombination energies between potential and carriers, respectively. The recombination also leads to the moving of potentials.

## III. Ground State

When temperature is very close to absolute zero, all carriers are paired and each potential is occupied by a pair of charges. Therefore, there are no unpaired carriers in high-T$_c$ system at T = 0 k. In this case, the above Hamiltonian can be simplified by removing the last term in Eq. (1).

$$H = \varepsilon_a \sum_i a_i^+ a_i + \sum_{<i,j>} v^a_{ij} n^a_i n^a_j + \varepsilon_b \sum_k b_k^+ b_k + \sum_{<k,l>} v^c_{kl} n^c_k n^c_l + \sum_{<i,k>} v^{ac}_{ik} n^a_i n^c_k + t_{ab} \sum_{<i,k>} (a_i^+ b_k^+ + hc.) \quad (2)$$

Let us consider a system doped with holes, where the potential is composed of two electrons. Therefore, $d_i^+$ and $d_i$ are the operators for the creation and annihilation of electron in potential, and $c_i^+$ and $c_i$ are the operators for creating and annihilating hole. In light of Hartree's approach, the wave function of basis state of superconductor is given by

$$|\varphi> = \prod(U_k + V_k S_k^+)|0> \quad (3)$$

$$S_k^+ = c_{k\uparrow}^+ c_{-k\downarrow}^+ \quad (4)$$

$$U_i^2 + V_i^2 = 1 \quad (5)$$

where, |0> is the vacant state. $S_k^+$ is the creation operator for a pair of electrons. $U_k^2$ and $V_k^2$ are the probability without and with a pair, respectively.

When Eq. (2) is multiplied by $\langle\varphi|$ from the left side of the Hamiltonian and $|\varphi\rangle$ from the right side of the equation, the energy (W) can be obtained as below

$$W = \sum_i \varepsilon_i^a V_i^2 + v^a \sum_{i,j} U_i^2 V_i^2 U_j^2 V_j^2 + \sum_k \varepsilon_k^b V_k^2 + v^c \sum_{k,l} U_l V_l U_k V_k$$
$$+ v^{ac} \sum_{i,k} U_i^2 V_i^2 U_k V_k + 2t_{ab} \sum_{i,k} U_i V_i U_k V_k \quad (6)$$

Let us define following parameters:

$$h_i = V_i^2, \quad 1 - h_i = U_i^2 \quad (7)$$

Then, W can be re-written as following:

$$W = \sum_i \varepsilon_i^a h_i + v^a \sum_{i,j} h_i(1-h_i)h_j(1-h_j) + \sum_k \varepsilon_k^b h_k$$
$$+ v^c \sum_{k,l} [h_k(1-h_k)h_l(1-h_l)]^{1/2} + v^{ac} \sum_{i,k} h_i(1-h_i)[h_k(1-h_k)]^{1/2}$$
$$+ 2t_{ab} \sum_{i,k} [h_i(1-h_i)h_k(1-h_k)]^{1/2} \quad (8)$$

Differentiating Eq. (8) to $h_i$ gives

$$\frac{\partial W}{\partial h_k} = \varepsilon_k^b - \Delta_0 \frac{1-2h_k}{2[h_k(1-h_k)]^{1/2}} \quad (9)$$

where, $\Delta_0$ is defined as

$$\Delta_0 = -\{v^c \sum_l [h_l(1-h_l)]^{\frac{1}{2}} + v^{ac} \sum_i h_i(1-h_i)$$
$$+ 2t_{ab} \sum_i [h_i(1-h_i)]^{\frac{1}{2}}\} \quad (10)$$

According to the principle of energy minimum, I consider

$$\frac{\partial W}{\partial h_k} = 0 \qquad (11)$$

Then, we have

$$h_k = \frac{1}{2}\left(1 - \frac{\varepsilon_k^b}{(\epsilon_k^{b^2}+\Delta_0^2)^{1/2}}\right) \qquad (12)$$

Then,

$$1 - h_k = \frac{1}{2}\left(1 + \frac{\varepsilon_k^b}{(\epsilon_k^{b^2}+\Delta_0^2)^{1/2}}\right) \qquad (13)$$

**IV. Excited State**

When T>0 K, Fermi function ($f_k$) has to be considered because there exist thermally excited particles in the system. In terms of Eq. (2) and above analysis to the system at T=0 K, we can obtain following formula:

$$W = \sum_i \varepsilon_i^a [f_i + (1-2f_i)h_i] + v^a \sum_{i,j} h_i(1-h_i)h_j(1-h_j)(1-2f_i)^2(1-2f_j)^2$$

$$+ \sum_k \varepsilon_k^b [f_k + (1-2f_k)h_k]$$

$$+ v^c \sum_{k,l} [h_k(1-h_k)h_l(1-h_l)]^{1/2}(1-2f_k)(1-2f_l)$$

$$+ 2t_{ab} \sum_{i,k} [h_i(1-h_i)h_k(1-h_k)]^{1/2}(1-2f_i)(1-2f_k)$$

$$+ v^{ac} \sum_{i,k} h_i(1-h_i)[h_k(1-h_k)]^{1/2}(1-2f_i)^2(1-2f_k)$$

$$+ 2t_{dc} \sum_{i,j,k} h_i(1-h_i)[h_j(1-h_j)h_k(1-h_k)]^{1/2}(1-2f_i)^2(1-2f_j)(1$$

$$- 2f_k)$$

$$(14)$$

The entropy of the system is defined as below:

$$TS = -k_B T \sum_k [f_k \ln f_k + (1-f_k)\ln(1-f_k)] \quad (15)$$

where $k_B$ is the Boltzmann constant. The free energy of the system is defined as F (=W+TS). When the system stays at a stable state, F must be minimal regarding to $h_k$ and $f_k$. According to this principle, we come to following minimal F with respect to $h_k$:

$$\frac{\partial F}{\partial h_k} = \varepsilon_k^b - \Delta_0 \frac{1-2h_k}{2[h_k(1-h_k)]^{\frac{1}{2}}} = 0 \quad (16)$$

where $\Delta_0$ is redefined as:

$$\Delta_0 = -\left\{ v^c \sum_l [h_l(1-h_l)]^{\frac{1}{2}}(1-2f_l) + 2t_{ab}\sum_i [h_i(1-h_i)]^{\frac{1}{2}}(1-2f_i) \right.$$

$$+ v^{ac}\sum_i h_i(1-h_i)(1-2f_i)^2$$

$$\left. + 2t_{dc}\sum_{i,j} h_i(1-h_i)[h_j(1-h_j)]^{\frac{1}{2}}(1-2f_i)^2(1-2f_j) \right\} \quad (17)$$

We can obtain:

$$h_k = \frac{1}{2}\left[1 - \frac{\varepsilon_k}{\left(\varepsilon_k^{b^2} + \Delta_0^2\right)^{\frac{1}{2}}}\right] \quad (18)$$

By defining $E_k = \left(\varepsilon_k^{b^2} + \Delta_0^2\right)^{\frac{1}{2}}$, we have $h_k = \frac{1}{2}\left(1 - \frac{\varepsilon_k}{E_k}\right)$

Similarly, we have zero F with respect to $f_k$ as following:

$$\frac{\partial F}{\partial f_k} = \varepsilon_k^b(1-2h_k) + 2\Delta_0[h_k(1-h_k)]^{\frac{1}{2}} + k_B T \ln\frac{f_k}{1-f_k} = 0 \quad (19)$$

Thus, we have:

$$f_k = \frac{1}{e^{\beta E_k} + 1}, \quad \beta = \frac{1}{k_B T} \tag{20}$$

For simplicity, we define parts in Eq. (18) as $\Delta_1$ and consider it is a constant

$$-\left\{2t_{ab}\sum_i [h_i(1-h_i)]^{\frac{1}{2}}(1-2f_i) + v^{ac}\sum_i h_i(1-h_i)(1-2f_i)^2 \right.$$

$$\left. + 2t_{dc}\sum_{i,j} h_i(1-h_i)[h_j(1-h_j)]^{\frac{1}{2}}(1-2f_i)^2(1-2f_j)\right\} = \Delta_1 \tag{21}$$

Then, we have

$$-v^c \sum_l [h_l(1-h_l)]^{\frac{1}{2}}(1-2f_l) = \Delta_0 - \Delta_1 \tag{22}$$

To get the expression of T$_c$, we define $V = -v^c$ and $\Delta = \Delta_0 - \Delta_1$. Substituting Eq. (18) and (20) into Eq. (21), the following answer is obtained

$$V\sum_l \frac{\Delta_0}{E_k} \times \frac{1-e^{\beta E_k}}{1+e^{\beta E_k}} = \Delta \tag{23}$$

The density of states at the Fermi level is given by

$$\frac{dN(E_k)}{dE_k} = \frac{dN(\varepsilon_k)}{d\varepsilon_k} \times \frac{d\varepsilon_k}{dE_k} = N_0 \frac{E_k}{\sqrt{E_k^2 - \Delta_0^2}} \tag{25}$$

where N(0) is the state density near the Fermi surface. When T ≤ Tc, it is considered that the total number of states keeps unchanged. Combining Eq. (23) and (24) and converting the summation into integration, we have

$$\frac{\Delta}{N_0 V \Delta_0} = \int_0^{\hbar\omega_c} \frac{d\varepsilon_k}{\sqrt{E_k^2 - \Delta_0^2}} \times \frac{1-e^{\beta E_k}}{1+e^{\beta E_k}} \times \frac{dE_k}{d\varepsilon_k}$$

$$= \int_0^{\hbar\omega_c} \frac{d\varepsilon_k}{\sqrt{\varepsilon_k^2 + \Delta_0^2}} \times \tanh[\frac{1}{2}\beta(\varepsilon_k^2 + \Delta_0^2)^{\frac{1}{2}}] \tag{25}$$

where $\omega_c$ is the highest vibrational frequency of localized charges (potential). After resolving the Eq. (25), the following appropriate result is attained

$$k_B T_c = A * \hbar\omega_c e^{-\frac{\Delta}{N_0 V \Delta_0}} = A * \hbar\omega_c e^{-\frac{1}{N_0 V}(1-\frac{\Delta_1}{\Delta_0})} \qquad (26)$$

A is a constant. Clearly, we see that the formula for $T_c$ is similar to that in the BCS theory (Eq. 3.29 in Ref. 9). The difference is that the interactions here are induced by the localized charges (local potentials) and mobile carriers. In Eq. (26), the frequency is related to the vibration of localized charges (or local potentials). At the same time, $\Delta$ and $\Delta_0$ play very important roles on the value of $T_c$, that is, the stronger the interactions between potentials and between potential and carrier, the higher $T_c$.

In above discussion, we focused on that the charge states of potential and carrier are opposite. If they are same, such as electron-doped cuprates, the last two items in Eq. (1) need to be removed. Then, Eq. (21) shall be:

$$\Delta_1 = -v^{ac} \sum_i h_i(1-h_i)(1-2f_i)^2 \qquad (27)$$

As the last two terms describe the recombination between the potential and carrier or the movement of potentials, their contribution to superconductivity also disappear, leading to relatively lower $T_c$. Therefore, the transition temperature for electron-doped system (Eq. (26)) should be lower than that of hole-doped ones.

**V. Discussion**

As indicated by the electron phase diagram of the HTS system, there exists the HTS state only if the carrier's density keeps within a range. The rich phase diagram is originated from the synergistic effect among various interactions. In this model, it is considered that there are two kinds of carriers, mobile one and localized one, in HTS. The localized carriers result in ordered potentials, which are analogous to the ions and mediate the paring of mobile carriers. At the same time, the potentials are movable by interacting with mobile carrier through recombination, contributing to the high transition temperature in HTS. If there is a paucity of carriers, the potentials are too localized to move/vibrate, and the carriers are too few to pair. When there are excessive carriers, the strength and ordering of potential are reduced and broken, respectively, which weakens the correlative vibration of potentials and various interactions. Therefore, only the optimal carrier density leads to high $T_c$, especially if the charge states local potential and carrier are opposite because of the strong interaction between them. Additionally, our Hamiltonian may provide a theoretical model to describe the psuedogap and strange metal behaviors as induced by the interactions at various carrier densities under certain temperature.

## VI. Conclusions

In summary, a carrier-potential interaction model is presented to describe the high-temperature superconducting behavior. We show that the potential is composed of localized charges, which vibrates and can be movable even. The Hamiltonian of the HTS state is obtained by analyzing various interactions. We find that the transition temperature is strongly correlated to the vibration of local potentials and the interactions among potentials and carriers, which result in high $T_c$ under the optimal carrier density. We further demonstrate that the formula to express the transition temperature based on the carrier-potential interaction is similar to that based on phonon-mediation. By replacing the local-potential with lattice point, it is the BCS model. Therefore, the origin of superconductivity is attributed to the pairing of carriers mediated by the vibration of periodic local potentials: (1) for BCS theory, phonon-mediation (lattice vibration) is dominant for low-$T_c$ superconductivity and (2) in our model, the carrier-potential interaction is responsible for high-$T_c$ superconductivity. Furthermore, our model may explain the mysterious phenomena of HTS observed at certain temperature.